Correspondence and requests for materials should be addressed to jianwangphysics@pku.edu.cn (Jian Wang) and kehe@aphy.iphy.ac.cn (Ke He)

# Crossover between Weak Antilocalization and Weak Localization of Bulk States in Ultrathin Bi$_2$Se$_3$ Films

Huichao Wang[1,2], Haiwen Liu[1,2], Cui-Zu Chang[3,4], Huakun Zuo[5], Yanfei Zhao[1,2], Yi Sun[1,2], Zhengcai Xia[5], Ke He[2,3,4,*], Xucun Ma[2,3,4], X. C. Xie[1,2], Qi-Kun Xue[2,3] and Jian Wang[1,2,*]

[1]International Center for Quantum Materials, School of Physics, Peking University, Beijing 100871, People's Republic of China

[2]Collaborative Innovation Center of Quantum Matter, Beijing, People's Republic of China

[3]State Key Laboratory of Low-Dimensional Quantum Physics, Department of Physics, Tsinghua University, Beijing 100084, People's Republic of China

[4]Institute of Physics, Chinese Academy of Sciences, Beijing 100190, People's Republic of China

[5]Wuhan National High Magnetic Field Center, Huazhong University of Science and Technology, Wuhan 430074, People's Republic of China

**ABSTRACT**

We report transport studies on the 5 nm thick Bi$_2$Se$_3$ topological insulator films which are grown via molecular beam epitaxy technique. The angle-resolved photoemission spectroscopy data show that the Fermi level of the system lies in the bulk conduction band above the Dirac point, suggesting important contribution of bulk states to the transport results. In particular, the crossover from weak antilocalization to weak localization in the bulk states is observed in the parallel magnetic field measurements up to 50 Tesla. The measured magneto-resistance exhibits interesting anisotropy with respect to the orientation of $B_{//}$ and I, signifying intrinsic spin-orbit coupling in the Bi$_2$Se$_3$ films. Our work directly shows the crossover of quantum interference effect in the bulk states from weak antilocalization to weak localization. It presents an important step toward a better understanding of the existing three-dimensional topological insulators and the potential applications of nano-scale topological insulator devices.

Topological insulators (TIs), a new type of quantum materials characterized by the existence of a gap in the energy spectrum and the symmetry-protected surface states within the bulk gap, have attracted significant interest in recent years[1-3]. As a result of the strong spin-orbit coupling (SOC), the spin-momentum locked surface states always show weak antilocalization (WAL) effect[4-6] due to the acquired π Berry phase when completing a time-reversed self-crossing path. Plentiful interesting behaviors of TIs, particularly in the topologically protected surface states, have been explored through electrical transport experiments[7-22]. Whereas, it is known that the bulk states of the existing three-dimensional (3D) TIs are not perfectly insulating even at low temperatures, and as a direct consequence, the bulk states can also influence the transport behaviors of TIs. In order to develop a comprehensive understanding on the exotic properties of TIs, it is important to systematically study the transport of TIs taking into account of the effects from the bulk states.

Contrasting to the symmetry protected surface states, the TI bulk states with strong SOC can lead to either WAL or weak localization (WL) depending on the experimental conditions[5,6,23]. Recently, the WL effect of bulk states has been studied in the ultrathin $Bi_2Se_3$ films under perpendicular magnetic field[24]. However, it is difficult to distinguish the bulk state features from the surface state behaviors by the perpendicular field transport measurements. In comparison, the parallel field MR properties of pure TI films reveal more bulk-state information[25,26], and therefore, the parallel magnetic field transport measurements provide a powerful tool in investigating the quantum interference effects in the TI bulk states.

Here we comprehensively study the transport properties of 5 quintuple layers (QLs) $Bi_2Se_3$ films grown on epi-ready $\alpha$-$Al_2O_3$ (0001) substrates by molecular beam epitaxy (MBE). The *in situ* angle-resolved photoemission spectroscopy (ARPES) result shows that two quantum well states of the conduction band, as well as the surface states, contribute to the density of states near the Fermi level. The MR is measured in pulsed high magnetic field (PHMF) up to 50 Tesla (T). When the external field is perpendicular to the film plane, linear MR is observed above 16 T. In parallel magnetic field, MR behaviors of the TI films exhibit interesting anisotropy when the field orientation is tuned between $B_{//} \perp I$ and $B_{//} // I$ configurations. Specifically, while the field is increased up to 50 T, the MR always exhibits a distinctive switch from positive to negative when $B_{//} \perp I$; whereas, when $B_{//} // I$, such switch either occurs at a much larger field, or does not occur at all within 50 T. The theoretical analysis and the quantitative fitting for the experimental results indicate that such parallel field MR behavior of the ultrathin TI films can be well explained using the WAL-WL crossover mechanism in the TI bulk states. To be specific, as a consequence of the strong SOC effect in the TI films, the phase coherence time $\tau_\varphi$ generally presents a larger time scale than the intrinsic spin-flip time $\tau_{SO}$. As a result, the WAL effect always occurs when the magnetic field is relatively weak, whereas, the WL effect arises when the field becomes sufficiently strong. From this perspective, the MR switch from positive to negative can be understood as a manifestation of the WAL-WL crossover in the TI bulk states. In addition, we observe novel MR anisotropy with respect to the relative orientation of the parallel magnetic field and the current, which can be qualitatively explained using the SOC mechanism. Combining the ARPES data, the transport results and the theoretical analysis, we reveal the crossover of quantum interference effect, from WAL to WL, in the bulk states of the ultrathin $Bi_2Se_3$ films.

## Results

The 5 QLs $Bi_2Se_3$ films are epitaxially grown on sapphire (0001) substrates in an ultra-high vacuum MBE-ARPES-STM (scanning tunneling microscope) combined system[27]. Figure 1(a) shows the reflection high-energy electron diffraction (RHEED) pattern of a typical MBE-grown 5 QLs $Bi_2Se_3$ film, where the sharp 1×1 streaks demonstrate that the film has good single crystal quality. The band map of the film taken in Γ–K direction is shown in Figure 1(b). In the ARPES measurement, photoelectrons are excited by an unpolarized He-I light (21.21 eV), and collected by a Scienta SES-2002 analyzer (15 meV). The topological surface states are observed within the bulk gap. A quite small gap is opened in the Dirac cone just below the Dirac point due to the coupling between the top and bottom surface states. Moreover, as shown in Figures 1(b) and (c), the quantum well states of the conduction band have a significant contribution to the Fermi surface. These observations indicate that our films are TIs with the Dirac surface states and the bulk states play a great role in the charge transport.

In order to explore the transport properties of the films by *ex situ* measurements, 20-nm-thick insulating amorphous Se is deposited on the TI films as a protection capping layer (Figure 1(d)). The Hall bars with channel dimensions of 600 × 400 $\mu m^2$ are fabricated using standard photo-lithography method. An optical image of the typical Hall bar structure is shown in the inset of Figure 2(a). The transport properties of the films were measured mainly by using the standard four-electrode method in a 16 T Quantum Design PPMS (physical property measurement system) and PHMF at the Wuhan National High Magnetic Field Center. The non-destructive pulsed magnet has a rise time of 15 ms and fall time of 135 ms. The field and current directions are reversed in the measurements to improve the precision of the results by eliminating the parasitic voltages. Typical results from the sample 1 (a 5 QLs $Bi_2Se_3$ film) are shown in the main text. (The results are repeatable in a collection of samples.)

The dependence of the sheet longitudinal resistance $R_☐$ on the temperature was measured firstly (for the sample 1), which is shown in Figure 2(a). The resistance decreases almost linearly when the temperature decreases from 300 K to about 50 K. Whereas, when the temperature drops approximately below 12 K, the resistance develops an upturn. The upturn resistance, as shown in Figure 2(b), exhibits a logarithmic dependence on the temperature, which can be understood in terms of the quantum correction induced by the electron-electron interaction[13]. The resistance upturn becomes larger at 8 T than that at 0 T, which is due to the suppression of the WAL effect in the presence of the perpendicular magnetic field resulting in enhanced resistance. The Hall trace of the sample 1 at 4.2 K is shown in Figure 2(c). Utilizing the slope at low fields, we calculate the sheet carrier density $n_s \sim 1.89 \times 10^{13}$ cm$^{-2}$ and mobility $\mu \sim 316.6$ cm$^2$/(V·s) at 4.2 K.

Next, the MR behavior of the sample 1 at relatively low perpendicular fields (< 16 T) was measured in PPMS and PHMF, respectively. The results in both measurements are consistent, which are shown in Figure 2(d). Specifically, these MR curves show typical dips around 0 T at low temperatures as a result of the WAL effect in TIs[7, 12-15], which can be clearly seen in the inset of Figure 2(d). In addition, a linear MR appears above 16 T at both 4.2 K and 77 K. This is consistent with the previous observation in a 5 QLs $Bi_2Se_3$ film grown on the same substrate by MBE[15]. Note that similar linear behavior of MR has been observed in many TI systems[28-35], yet, the underlying origin remains to be clarified. Some attribute such linear MR to a quantum origin, including the Abrikosov's quantum linear MR model which relates the linear MR to the linear energy dispersion of the gapless topological surface states[28-30], and the Hikami-Larkin-Nagaoka theory which interprets the linear-like MR at high fields in terms of the WAL mechanism[31,32]. Other theories associating the linear MR with a classical origin have also been proposed, such as the inhomogeneity model by Parish and Littlewood which stresses the influence from the sample inhomogeneity[34,35].

Interestingly, a different set of rich MR behavior is observed when a magnetic field parallel to the film plane is applied. In particular, as clearly shown in Figure 3, the normalized MR of sample 1 along the [110] direction in parallel field, exhibits a significant dependence on the relative orientation of $B_{//}$ and I. The behavior of MR when the parallel field is oriented along $B_{//} \perp I$ (the field increases up to 50 T) is shown in Figure 3(a). At 4.2 K, the MR first increases to a maximum at about 25 T, and then decreases as a downward parabolic-like curve. When the temperature is increased to 77 K, the MR slightly increases within 11 T and then decreases monotonically up to 50 T. In comparison, the MR behavior for the field orientation $B_{//} // I$ (Figure 3(b)) is different, only showing positive MR within 50 T at both 4.2 K and 77 K.

To further understand the parallel field MR behavior, we also measured the parallel field MR along the [-110] direction of the sample 1. The results are shown in Figures 4(a) and (b). The quasi-four-electrode method (insets of Figures 4(a) and (b)) is used only here due to the restriction of sample size. It is worth mentioning that such measurement method can introduce contact resistance that can lead to a weaker MR response without changing the curve shape (Supplementary Information). In this case, the MR switch from positive to negative is also clearly observed at 4.2 K and 77 K when $B_{//} \perp I$, but much weaker when $B_{//} // I$. For a fixed temperature, the switching field for $B_{//} // I$ is larger than that for $B_{//} \perp I$. In order to explore the general dependence of the MR switch on the $B_{//}$-I orientation, the parallel field MR of sample 1 with the current along the [110] direction under different position $\theta$ was measured (Figure 4(c)). The sample position $\theta$ refers to the angle between $B_{//}$ and the normal direction of I along the film plane (the inset of Figure 4(c)). Here $\theta=0°$ means $B_{//} \perp I$ and $\theta=90°$ means $B_{//} // I$. When the magnetic field increases up to 16 T, the MR at 100 K increases when $B_{//} // I$, but decreases when $B_{//} \perp I$. When $\theta=45°$, the MR seems to show a competition between positive and negative MR. Normalized MR at 100 K as a function of $\theta$ when $B_{//}$ = 15 T is shown in Figure 4(d).

## Discussion

A positive-to-negative MR switch of the ultrathin $Bi_2Se_3$ films is observed when the parallel magnetic field increases. The switch occurs in a broad range of temperature and exhibits anisotropy when the relative orientation of $B_{//}$ and I is varied. The parallel field MR data exhibit tiny variation in the presence of a large magnetic field, indicating a quantum correction regime. Taking into account the fact that the Fermi level locates at the bulk conduction band, and that the film shows quite large carrier density and lower mobility compared to those reported for the surface states of TI[10], we conclude that the major contribution to the charge transport in our films comes from the bulk states instead of the surface states. We emphasize that, while similar negative MR behavior observed in Sn-doping $Bi_2Te_3$ films[26] under parallel field was reported, a clear interpretation of the MR data and the comparison of different behaviors in Sn-doping samples are still missing. In contrast, for our present experimental results, the parallel field MR switch phenomenon of $Bi_2Se_3$ films can be well explained in terms of the quantum interference effects in the TI bulk states. In particular, as we shall describe in detail below, the positive and negative MR can be attributed to the presence of WAL and WL, respectively.

For a TI system with strong SOC, the correction to the MR behavior due to the effect of quantum interference[6] is characterized by two time scales: the dephasing time $\tau_\varphi$ and the spin-flip time $\tau_{SO}$. In the regime $\tau_\varphi \gg \tau_{SO}$, the spin-orbit scattering is significant, which gives rise to frequent spin flips. As a result, the quantum interference between the time reversed trajectories is destructive, leading to the occurrence of WAL and a positive MR. In the regime where $\tau_\varphi$ is comparable to $\tau_{SO}$ or $\tau_\varphi < \tau_{SO}$, the spin-orbit scattering effect is weak and the scattering process barely affects the spin orientation. In such case, negative MR occurs with increasing magnetic field as a WL feature. In the intermediate regime where $\tau_\varphi > \tau_{SO}$, MR first increases to a peak value and then decreases, signifying a crossover from WAL to WL. Overall, the quantum interference behavior of a system is crucially dependent on the ratio between $\tau_\varphi$ and $\tau_{SO}$.

In our thin films, the mean free path $l$ can be estimated according to $l = \hbar k_F \mu / e$, where $k_F$ can be obtained through the sheet density $n_s$. For the sample 1 described in this letter, $l$ is approximately 20 nm, much larger than the thickness $d = 5$ nm. The parallel field transport thereby falls into the Dugaev-Khmelnitskii (DK) regime where the resistance change is given by[36,37]

$$\frac{\Delta R}{R} = -N_i \frac{e^2 R_\square}{2\pi^2 \hbar} \left[ \frac{3}{2} \ln\left(1 + \frac{L_{so}^2}{L_\parallel^2}\right) - \frac{1}{2} \ln\left(1 + \frac{L_\varphi^2}{L_\parallel^2}\right) \right]. \quad (1)$$

Here $\Delta R = R(B) - R(B=0)$. R can be regarded as the resistance in the absence of external magnetic field due to the quite small variation, and $R_\square$ is the sheet resistance. $N_i$ is the number of channels contributing to the charge transport. In addition, three length scales are involved: $L_{so}$ is the spin-flip length (the distance travelled by an electron before its spin direction is changed by the scattering, which depends on the strength of SOC in the bulk states. Please see Supplementary Information.), $L_\varphi$ is the phase coherence length, and $L_\parallel = 4L_B^2\sqrt{l/d}/d$ with $L_B$ being the magnetic length $L_B = \sqrt{\hbar/eB}$. By varying only two fitting parameters $L_\varphi$ and $L_{SO}$, we present a quantitative fitting of the experimental data in Figure 3. In the case of $B_{//} \perp I$, we have $L_\varphi \approx 140$ nm, $L_{SO} \approx 30$ nm at 4.2 K and $L_\varphi \approx 40$ nm, $L_{SO} \approx 20$ nm at 77 K. And for $B_{//} // I$, $L_\varphi \approx 140$ nm, $L_{SO} \approx 18$ nm at 4.2 K. Based on these fitting results, $L_\varphi \approx 40$ nm and $L_{SO} \approx 14$ nm at 77 K when $B_{//} // I$ can be obtained by calculation (see Supplementary Information for details). Substituting these calculated values into Equation (1), we quantitatively reproduce the MR result at 77 K in $B_{//} // I$ configuration as shown in Figure 3(b). The tiny mismatches of the fitting MR curves may arise from the neglect of the electron-electron interaction[13] here, which can actually play a role in the MR behavior.

It is noted that $N_i=2$ in the fitting is identical to the number of bulk bands at the Fermi surface shown in Figure 2(b). While we cannot fully determine whether the two channels are completely from the two subbands of the bulk state or from the hybridized channels of bulk state and surface state, it is evident that the bulk state plays dominant role in the emergence of crossover behavior reported here. All the values of parameters shown above are fitting results and reasonable. The phase-coherence length $L_\varphi$ decreases with temperature and $L_\varphi$ is much larger than the mean free path as reported earlier[15]. Besides, the frequent scattering due to the edges in the ultrathin films makes $L_{so}$ comparable to the mean free path. Based on the fitting results of experimental data, $L_{SO}$ is affected by the relative orientation of the parallel magnetic field and the current.

Through the parallel field MR curves of the ultrathin $Bi_2Se_3$ films, a clear image of the quantum interference effects with the WAL-WL crossover in the TI bulk states is depicted. Three cases are presented for the data in Figure 3 and 4. (I) For the $B_{//} \perp I$ configuration, the system usually falls into the intermediate regime ($\tau_\varphi > \tau_{SO}$) at relatively low temperatures. As a result, WAL effect characteristically arises when the field is weak, whereas, when the magnetic field becomes sufficiently large, a crossover into WL occurs. The data of Figure 3(a) and 4(a) belong to this case and the crossover is observed within 50 T. (II) For the $B_{//}$ // I configuration and relatively low temperatures, compared to (I), a weaker effect of the WAL-WL crossover is observed at a larger switching field due to a larger ratio $\tau_\varphi/\tau_{SO}$ as shown in Figure 4(b). In Figure 3(b), only the WAL effect is observed at 4.2 K and 77 K. This may indicate that the switching fields from WAL to WL in both conditions are larger than 50 T. (III) For either of the two geometries, the MR may show only WL effect at high temperatures when $\tau_\varphi$ decreases and becomes comparable to $\tau_{SO}$, even $\tau_\varphi < \tau_{SO}$. The $\theta=0°$ data at 100 K in Figure 4(c) fall into this situation.

The observed anisotropy feature of MR can be related to many factors, including the band structure asymmetry, crystalline anisotropy, and the relative orientation of the parallel magnetic field and the current. However, generally speaking, both the band asymmetry and the crystal anisotropy could result in a MR behavior that varies along different crystal directions. Since the anisotropy of MR along the [-110] direction remains consistent with that of the [110] direction, we consider the relative directions of the parallel magnetic field and the current as the leading factor that contributes to the novel MR anisotropy reported here. The underlying physics can then be qualitatively understood in terms of the SOC mechanism. Consider a moving electron, which experiences an effective magnetic field $B_e$ that is perpendicular to the electron momentum. When $B_{//} \perp I$, the main transport path of the electron is perpendicular to the external field, in which the spin-flip process due to the SOC effect is strongly suppressed by the external magnetic field, leading to an effectively larger $\tau_{SO}$. In distinct contrast, when $B_{//}$ // I, the spin-flip process is hardly affected. This anisotropy in the effective SOC effect provides a mechanism that leads to the observed anisotropic MR here. Such explanation is also supported by Figures 4(c) and (d), which show an intimate relation between the parallel field MR and the sample position $\theta$. Particularly, the negative MR behavior when $B_{//} \perp I$ is consistent with a larger effective $\tau_{SO}$, while the positive MR when $B_{//}$ // I agrees with a smaller $\tau_{SO}$. For arbitrary $\theta$, the effective $\tau_{SO}$ is given by the combination of both effects, which leads to the MR-$\theta$ characteristics shown in Figure 4(d).

Interestingly, the anisotropy of the parallel field MR in thicker TI films (e.g. 45 QLs and 200 QLs Bi$_2$Se$_3$)[38] seems to be opposite to the experimental results of the 5 QLs Bi$_2$Se$_3$ films. Additionally, there is one order of magnitude difference of the MR amplitudes for the thicker films[38] between B$_{//}$⊥I and B$_{//}$//I, while it is not the case for the results presented here. In light of many different factors between the thick and ultrathin films, such as the thickness and the weight of bulk states at the Fermi surface, we are not able to achieve a unified understanding. Further theoretical investigations in this context would be valuable towards a quantitative comprehension of the anisotropic parallel field MR of TI films in various field-current orientations.

In summary, the crossover from WAL to WL in TI bulk states is revealed clearly by the parallel field MR behaviors of the 5 QLs ultrathin Bi$_2$Se$_3$ films. The WAL-WL crossover is largely affected by the relative orientation of the parallel magnetic field and the current, which can be understood in terms of an orientation-dependent effective spin-flip time. To be specific, when B$_{//}$ ⊥ I, the MR within 50 T always exhibits the crossover at 4.2 K and 77 K, whereas, the crossover for B$_{//}$ // I occurs at a much higher magnetic field (possibly above 50 T), due to the anisotropic SOC effect under strong magnetic field. The novel transport properties of the ultrathin TI films demonstrated in this work are quite significant for a better understanding of the existing 3D TI materials and for the future potential magnetic applications of nano-scale TI devices.

**Methods**

**Sample preparation.** The high quality 5 QLs Bi$_2$Se$_3$ films studied here were grown on α-Al$_2$O$_3$ (sapphire) (0001) substrates in an ultra-high vacuum MBE-ARPES-STM combined system with a base pressure better than 2×10$^{-10}$ mbar. Before sample growth, the sapphire substrates were first outgassed at 650 °C for 90 min and then heated at 850 °C for 30 min. High-purity Bi (99.9999%) and Se (99.999%) were evaporated from standard Knudsen cells. To reduce Se vacancies in Bi$_2$Se$_3$, the growth was kept in Se-rich condition with the substrate temperature at ~220 °C. For the transport measurement, 20 nm Se was capped on the TI film as protection layer.

**Device fabrication and measurement.** Hall bar structure with channel dimensions of 600 x 400 μm$^2$ was fabricated using standard photo-lithography method. Positive photoresist S1813 was spun at 4000 rpm for 45 seconds on Bi$_2$Se$_3$ film, followed by 110 °C baking for 60 seconds. With a mask of Hall bar pattern, the photoresist-coated sample was exposed to ultraviolet light (365nm wavelength) with exposure power of 8mW/cm$^2$ for 7 seconds. The exposed part of photoresist was removed after 40 seconds of developing (MFCD-26 developer). Then, bared area of Bi$_2$Se$_3$ film with no photoresist on was wet-etched with 1 g of potassium dichromate in 10 ml of sulfuric acid and 20 ml of DI water. The expected etch rate for Bi$_2$Se$_3$ was about 60 nm per minute, and DI water rinsing was needed. Via indium contacting, Au wires were attached as leads on samples at room temperature. Thus we were able to measure the transport properties of the fabricated devices by standard or quasi four-electrode method. Transport measurements were performed in PPMS (16T) and PHMF up to 50 T.

1. Qi, X.-L. & Zhang, S.-C. The quantum spin Hall effect and topological insulators. *Phys. Today* **63**, 33-38 (2010).
2. Hasan, M. Z. & Kane, C. L. Colloquium: Topological insulators. *Rev. Mod. Phys.* **82**, 3045-3067 (2010).
3. Qi, X.-L. & Zhang, S.-C. Topological insulators and superconductors. *Rev. Mod. Phys.* **83**, 1057-1110 (2011).
4. Lu, H.-Z., Shi, J. R. & Shen, S.-Q. Competition between weak localization and antilocalization in topological surface states. *Phys. Rev. Lett.* **107**, 076801 (2011).
5. Hikami, S., Larkin, A. I. & Nagaoka, Y. Spin-orbit interaction and magnetoresistnce in the two dimensional random system. *Prog. Theor. Phys.* **63**, 707-710 (1980).
6. Bergmann, G. Weak localization in thin films: a time-of-flight experiment with conduction electrons. *Phys. Rep.* **107**, 1-58 (1984).
7. Checkelsky, J. G. *et al.* Quantum interference in macroscopic crystals of nonmetallic Bi$_2$Se$_3$. *Phys. Rev. Lett.* **103**, 246601 (2009).
8. Taskin, A. A. & Ando, Y. Quantum oscillations in a topological insulator Bi$_{1-x}$Sb$_x$. *Phys. Rev. B* **80**, 085303 (2009).
9. Eto, K., Ren, Z., Taskin, A. A., Segawa, K. & Ando, Y. Angular-dependent oscillations of the magnetoresistance in Bi$_2$Se$_3$ due to the three-dimensional bulk Fermi surface. *Phys. Rev. B* **81**, 195309 (2010).
10. Qu, D.-X., Hor, Y. S., Xiong, J., Cava, R. J. & Ong, N. P. Quantum oscillations and Hall anomaly of surface states in the topological insulator Bi$_2$Te$_3$. *Science* **329**, 821-824 (2010).
11. Peng, H. *et al.* Aharonov–Bohm interference in topological insulator nanoribbons. *Nat. Mater.* **9**, 225-229 (2010).
12. Chen, J. *et al.* Gate-voltage control of chemical potential and weak antilocalization in Bi$_2$Se$_3$. *Phys. Rev. Lett.* **105**, 176602 (2010).

13. Wang, J. *et al.* Evidence for electron-electron interaction in topological insulator thin films. *Phys. Rev. B* **83**, 245438 (2011).
14. He, H.-T. *et al.* Impurity effect on weak antilocalization in the topological insulator $Bi_2Te_3$, *Phys. Rev. Lett.* **106**,166805 (2011).
15. Liu, M. H. *et al.* Electron interaction-driven insulating ground state in $Bi_2Se_3$ topological insulators in the two-dimensional limit. *Phys. Rev. B* **83**, 165440 (2011).
16. Steinberg, H., Gardner, D. R., Lee, Y. S. & Jarillo-Herrero, P. Surface state transport and ambipolar electric field effect in $Bi_2Se_3$ nanodevices. *Nano Lett.* **10**, 5032-5036 (2010).
17. Yuan, H. T. *et al.* Liquid-gated ambipolar transport in ultrathin films of a topological insulator $Bi_2Te_3$. *Nano Lett.* **11**, 2601-2605 (2011).
18. Wang, Y. *et al.* Gate-controlled surface conduction in Na-doped $Bi_2Te_3$ topological insulator nanoplates. *Nano Lett.* **12**, 1170-1175 (2012).
19. Fang, L. *et al.* Catalyst-free growth of millimeter-long topological insulator $Bi_2Se_3$ nanoribbons and the observation of the π-Berry phase. *Nano Lett.* **12**, 6164-6169 (2012).
20. Tian, M. L. *et al*. Dual evidence of surface Dirac states in thin cylindrical topological insulator $Bi_2Te_3$ nanowires. *Sci. Rep.* **3**, 1212 (2013).
21. Yan, Y. *et al.* Synthesis and quantum transport properties of $Bi_2Se_3$ topological insulator nanostructures. *Sci. Rep.* **3**, 1264 (2013).
22. Zhao, Y. F. *et al*. Demonstration of surface transport in a hybrid $Bi_2Se_3$/$Bi_2Te_3$ heterostructure. *Sci. Rep.* **3**, 3060 (2013).
23. Lu, H.-Z. & Shen, S.-Q. Weak localization of bulk channels in topological insulator thin films. *Phys. Rev. B* **84**, 125138 (2011).
24. Zhang, L. *et al*. Weak localization effects as evidence for bulk quantization in $Bi_2Se_3$ thin films. *Phys. Rev. B* **88**, 121103(R) (2013).
25. Lin, C. J. *et al.* Parallel field magnetoresistance in topological insulator thin films. *Phys. Rev. B* **88**, 041307(R) (2013).
26. Zhang, H. B. *et al.* Weak localization bulk state in a topological insulator $Bi_2Te_3$ film. *Phys. Rev. B* **86**, 075102 (2012).
27. Chang, C.-Z. *et al.* Growth of quantum well films of topological insulator $Bi_2Se_3$ on insulating substrate. *Spin* **1**, 21-25 (2011).
28. Tang, H., Liang, D., Qiu, R. L. J., & Gao, X. P. A. Two-dimensional transport-induced linear magneto-resistance in topological insulator $Bi_2Se_3$ nanoribbons. *ACS Nano.* **5**, 7510-7516 (2011).
29. He, H. T. *et al.* High-field linear magneto-resistance in topological insulator $Bi_2Se_3$ thin films. *Appl. Phys. Lett.* **100**, 032105 (2012).
30. Gao, B. F. Gehring, P., Burghard, M., and Kern, K. Gate-controlled linear magnetoresistance in thin $Bi_2Se_3$ sheets. *Appl. Phys. Lett.* **100**, 212402 (2012).

31. Zhang, S. X. *et al.* Magneto-resistance up to 60 Tesla in topological insulator $Bi_2Te_3$ thin films. *Appl. Phys. Lett.* **101**, 202403 (2012).
32. Assaf, B. A. *et al.* Linear magnetoresistance in topological insulators: Quantum phase coherence effects at high temperatures. *Appl. Phys. Lett.* **102**, 012102 (2013).
33. Wang, C. M. & Lei, X. L. Linear magnetoresistance on the topological surface. *Phys. Rev. B* **86**, 035442 (2012).
34. He, X. Y. *et al.* Highly tunable electron transport in epitaxial topological insulator $(Bi_{1-x}Sb_x)_2Te_3$ thin films. *Appl. Phys. Lett.* **101**, 123111 (2012).
35. Tian, J, *et al.* Quantum and classical magnetoresistance in ambipolar topological insulator transistors with gate-tunable bulk and surface conduction. *Sci. Rep.* **4**, 4859 (2014).
36. Altshuler, B. L.*,* Aronov, A. G., Khmelnitskii, D. E. & Larkin, A. I. in *Quantum Theory of Solids*, edited by Lifshitz, I. M. (Mir Publishers, Moscow, 1982).
37. Dugaev, V. K. & Khmelnitskii, D. E. Magnetoresistance of metallic films with low impurity concentration in a parallel magnetic field. *Sov. Phys. JETP* **59**,1038-1041 (1984).
38. Wang, J. *et al.* Anomalous anisotropic magnetoresistance in topological insulator films. *Nano Res.* **5**, 739-746 (2012).

**ACKNOWLEDGMENT**

We thank Liang Li for helpful discussions about the pulsed magnetic field measurements. This work was financially supported by the National Basic Research Program of China (Grant Nos. 2013CB934600 & 2012CB921300), the National Natural Science Foundation of China (Nos. 11222434 & 11174007), and the Research Fund for the Doctoral Program of Higher Education (RFDP) of China.

**Author contribution statement**

J.W. and K.H. conceived and designed the study. H.W., Y.Z., Y.S. and J.W. carried on the transport measurements. C.C. did MBE growth and ARPES experiment. H.Z. and Z.X. helped in pulsed high magnetic field measurements. K.H., X.M. and Q.X. supervised the MBE growth and ARPES experiment. H.W., C.C., J.W., H.L. and X.C.X. analyzed the data. H.L. and X.C.X. did theoretical fittings. H.W., H.L. and J.W. wrote the manuscript.

**Additional information**

**Supplementary information** accompanies this paper.

**Competing financial interests:** The authors declare no competing financial interests.

**Figure legends**

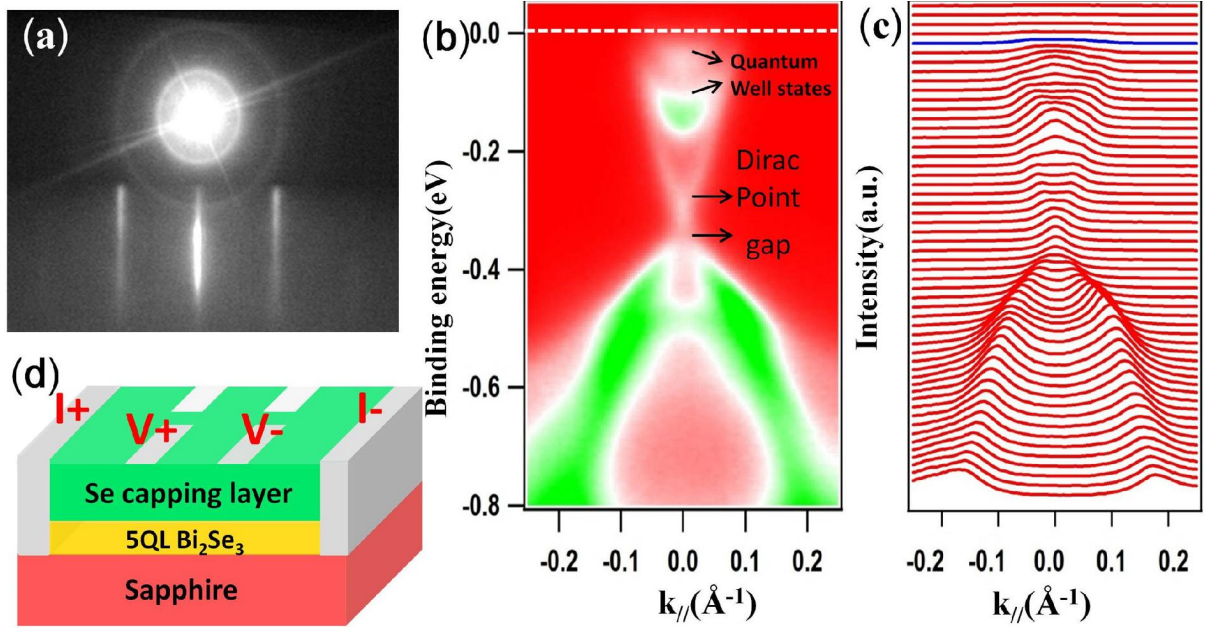

**Figure 1** *In situ* **characterizations of the 5QLs Bi$_2$Se$_3$ thin film grown by MBE and schematic structure for *ex situ* transport measurement.** (a) RHEED pattern with sharp 1×1 streaks. (b) ARPES band spectra along the Γ-K direction. The white dashed line indicates the Fermi level. The arrows label the quantum well states, the Dirac point and the small gap, respectively. (c) The corresponding momentum distribution curves of (b). The blue line indicates the Fermi level. (d) The schematic structure for transport measurements. The thickness is not to scale.

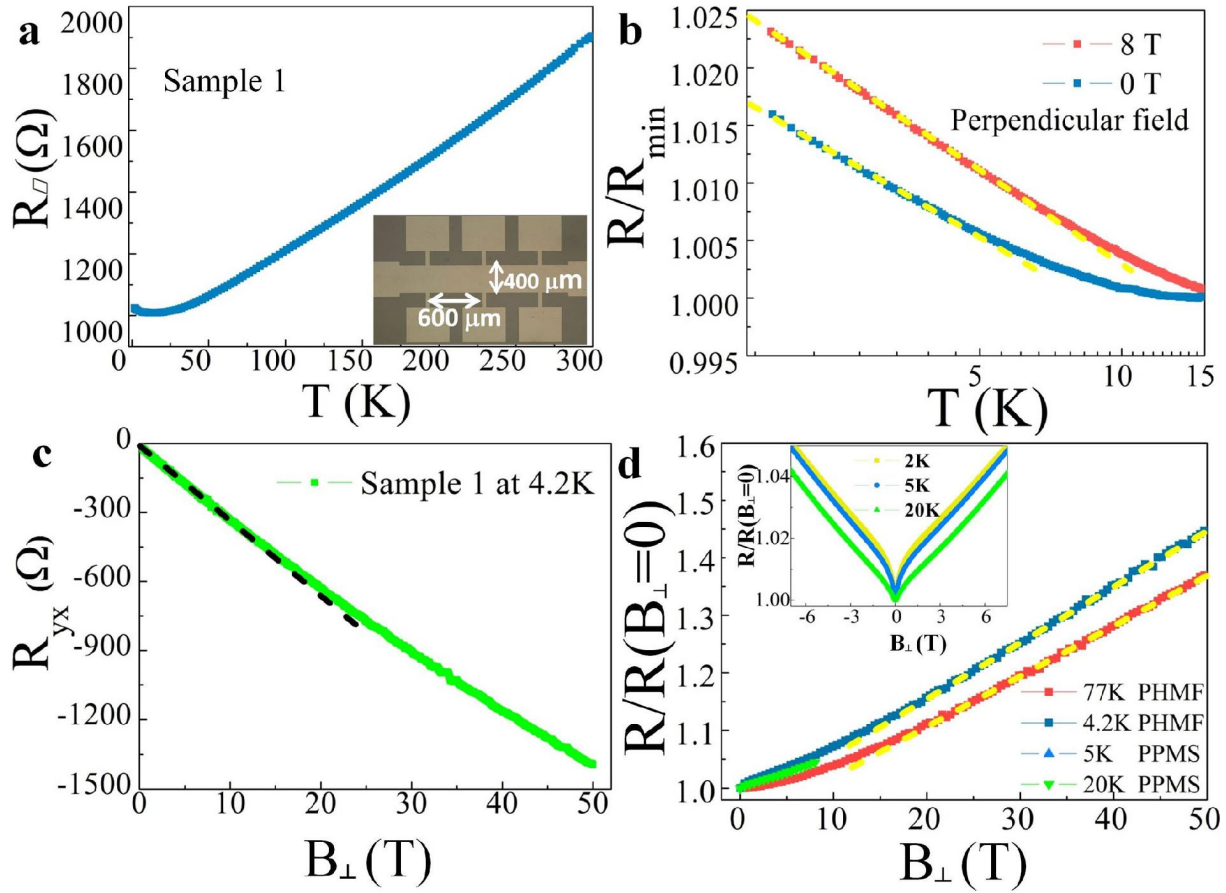

**Figure 2 Transport properties of sample 1.** (a) The sheet resistance $R_\square$ versus temperature with an upturn in low temperature regime. The inset is an optical image of the Hall bar structure. (b) Normalized upturn resistances at fixed perpendicular fields. (c) Hall trace at 4.2 K in a magnetic field up to 50 T. (d) Normalized perpendicular field MR. The inset shows the MR dips around 0 T which are attributed to the WAL effect at low temperatures. The broken lines are guides to the eyes.

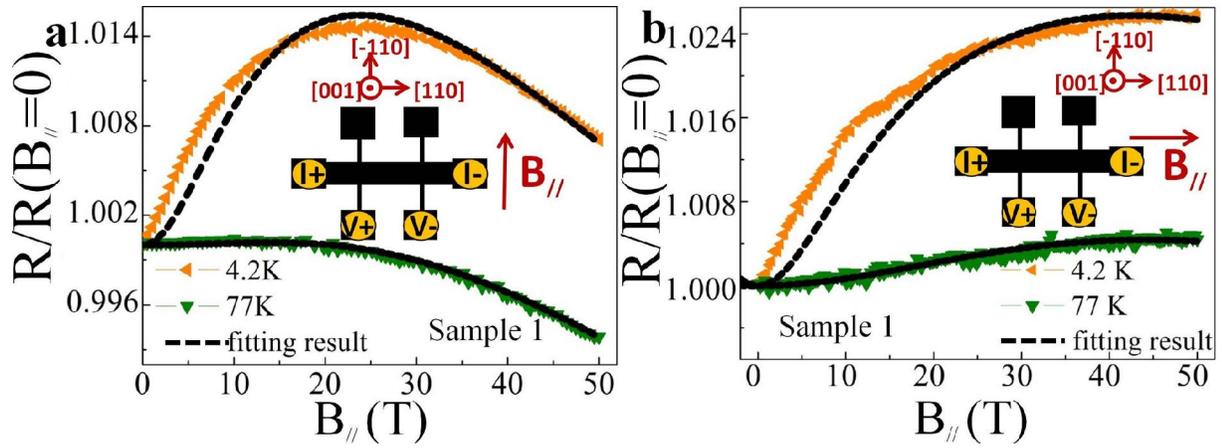

**Figure 3 Normalized MR of sample 1 in parallel field along [110] direction.** (a) $B_{//} \perp I$ and (b) $B_{//} // I$. The triangles are experimental data and the black dashed lines are fitting curves by Equation (1). The MR behaviors can be well explained by the WAL-WL crossover mechanism in TI bulk states. The insets show the corresponding measurement configurations. [110] and [-110] are two different crystal directions along the film plane.

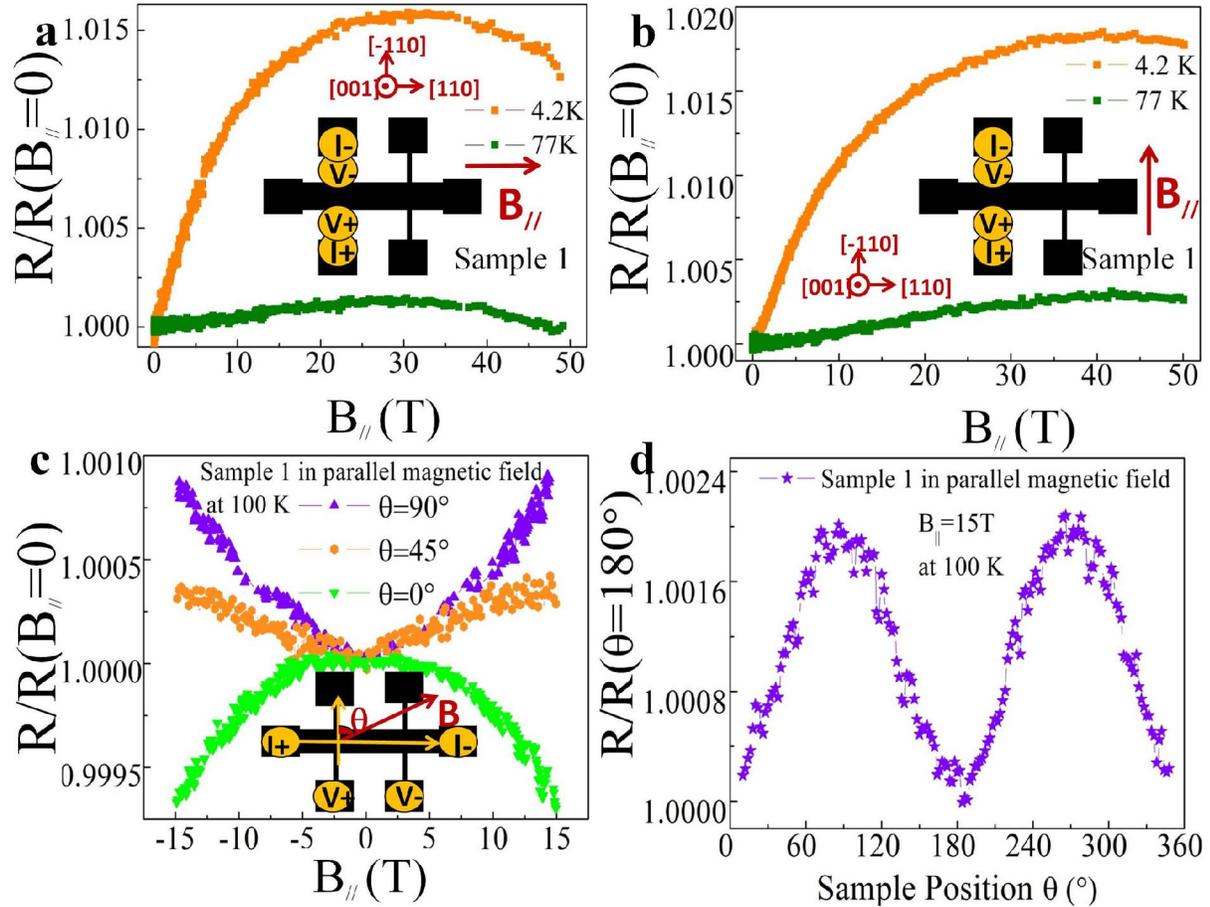

**Figure 4 Control experiments with sample 1.** Normalized MR along [-110] direction when (a) $B_{//} \perp I$ and (b) $B_{//} // I$. Normalized MR along [110] direction at 100 K when (c) θ=0°, 45°, 90° and (d) θ continuously varies from 0° to 90°. θ=0° means $B_{//} \perp I$ and θ=90° means $B_{//} // I$. The insets show the corresponding measurement configurations.

Supplementary Information for

"Crossover between Weak Antilocalization and Weak Localization of Bulk States in Ultrathin Bi$_2$Se$_3$ Films"

Huichao Wang[1,2], Haiwen Liu[1,2], Cui-Zu Chang[3,4], Huakun Zuo[5], Yanfei Zhao[1,2], Yi Sun[1,2], Zhengcai Xia[5], Ke He[2,3,4,*], Xucun Ma[2,3,4], X. C. Xie[1,2], Qi-Kun Xue[2,3] and Jian Wang[1,2,*]

[1]International Center for Quantum Materials, School of Physics, Peking University, Beijing 100871, People's Republic of China

[2]Collaborative Innovation Center of Quantum Matter, Beijing, People's Republic of China

[3]State Key Laboratory of Low-Dimensional Quantum Physics, Department of Physics, Tsinghua University, Beijing 100084, People's Republic of China

[4]Institute of Physics, Chinese Academy of Sciences, Beijing 100190, People's Republic of China

[5]Wuhan National High Magnetic Field Center, Huazhong University of Science and Technology, Wuhan 430074, People's Republic of China

* jianwangphysics@pku.edu.cn; kehe@iphy.ac.cn

I. Theoretical background

For the 5 quintuple layers Bi$_2$Se$_3$ topological insulator (TI) films, the surface states are weakly affected by the applied parallel magnetic field. For the parallel field magneto-resistance (MR) of the thin films, the main contribution is the phase factor $\exp\left(i\frac{e}{c\hbar}\int A dr\right)$ arising from the bulk states. The cooperon (particle-hole scattering diagram) in a system with strong spin-orbit coupling without magnetic field is [1,2]

$$C(q) = \frac{3}{2}\frac{1}{Dq^2 + \frac{1}{\tau_\varphi} + \frac{2}{\tau_{SO}}} - \frac{1}{2}\frac{1}{Dq^2 + \frac{1}{\tau_\varphi}}. \quad (S1)$$

Here, $q$ is the momentum, $D$ is the diffusion coefficient, $\tau_\varphi$ is the phase coherence time and $\tau_{SO}$ is the spin-flip time. By integrating all the cooperon terms, the correction to conductivity due to quantum interference is

$$\Delta\sigma = -\frac{2\sigma}{\pi v}C(r,r) = -\frac{2\sigma}{\pi v}\int\frac{d^2q}{(2\pi)^2}C(q). \quad (S2)$$

In the above, $\sigma$ is the conductivity and $v$ is the density of states. The mean free path $l$ in the 5 quintuple layers Bi$_2$Se$_3$ film is about 20 nm, much larger than the thickness d=5 nm. In this approximately $l \gg d$ limit, the diffusion equation for the cooperon in parallel magnetic field is given by

$$C(q) = \frac{3}{2}\frac{1}{Dq^2 + \frac{1}{\tau_B} + \frac{1}{\tau_\varphi} + \frac{2}{\tau_{SO}}} - \frac{1}{2}\frac{1}{Dq^2 + \frac{1}{\tau_B} + \frac{1}{\tau_\varphi}}. \quad (S3)$$

$\tau_B$ is defined by $\frac{1}{\tau_B} = \frac{1}{\tau}\frac{ld^3}{16L_B^4}$ where $\tau$ is the elastic scattering time and $L_B$ is the magnetic length with $L_B = \sqrt{\frac{\hbar}{eB}}$. We define a characteristic time $\tau_1^{-1} = \tau_\varphi^{-1} + 2\tau_{SO}^{-1}$ and the corresponding length in this time scale is $L_{SO} = \sqrt{D\tau_1}$ where D is the diffusion coefficient. Then the correction to the conductivity is given by [1,2]

$$\Delta\sigma = -N_i\frac{e^2}{2\pi^2\hbar}\left[\frac{3}{2}\ln\left(1+\frac{ld^3}{16L_B^4}\frac{\tau_1}{\tau}\right) + \frac{1}{2}\ln\left(1+\frac{ld^3}{16L_B^4}\frac{\tau_\varphi}{\tau}\right)\right]. \quad (S4)$$

Here N$_i$ is the number of energy bands contributing to the conductivity. By using the dephasing length $L_\varphi = \sqrt{D\tau_\varphi}$ and $L_{SO} = \sqrt{D\tau_1}$, we have the Equation (1) in the main text.

We introduce the detail calculation process for the parameters L$_\varphi$ and L$_{so}$ at 77 K when B$_{//}$ // I here. We assume the phase coherence length L$_\varphi$ is dependent only on temperature thus L$_\varphi$≈40 nm at 77 K when B$_{//}$ ⊥ I. Moreover, considering the different effective spin-flip time τ$_{SO}$ demonstrated by the anisotropic MR between B$_{//}$ // I and B$_{//}$ ⊥ I, we define and find a ratio of τ$_{SO}$ (B$_{//}$ // I) / τ$_{SO}$ (B$_{//}$ ⊥ I) from the 4.2 K fitting results. Utilizing the ratio and L$_{SO}$≈18 nm for B$_{//}$ ⊥ I at 77 K, L$_{SO}$≈14 nm for B$_{//}$ // I is deduced. The fitting curve utilizing L$_\varphi$≈40 nm and L$_{SO}$≈14 nm can quantitatively reproduce the MR behaviors at 77 K for B$_{//}$ // I configuration.

II. Transport properties of sample 2

Sample 2 is another 5 quintuple layers $Bi_2Se_3$ film on the sapphire (0001) substrate grown in the same MBE condition with sample 1. Similarly, 20 nm thick insulating amorphous Se is deposited on the film as capping layer for transport measurements. Figure S1(a) shows the sheet resistance-temperature characteristic of sample 2, which resembles that of sample 1 as shown in Fig. 2(a) of the main text. When applying a magnetic field perpendicular to the film plane, linear MR is observed at high fields, as shown in Fig. S1(b). In the parallel magnetic field, anisotropic MR behaviors along [110] direction are observed. With the increasing magnetic field when $B_{//}$ ⊥ I (Fig. S1(c)), the MR exhibits a switch from positive to negative at 4.2 K and 77 K with different switching fields, while only positive MR is observed within 50 T at both 4.2 K and 77 K when $B_{//}$ // I (Fig. S1(d)). These MR behaviors can be well explained by the WAL-WL crossover mechanism in TI bulk states as well. At the $B_{//}$ ⊥ I configuration, the MR switch at 4.2 K and 77 K corresponds to the crossover from WAL to WL. In the $B_{//}$ // I case, the positive MR indicates that only WAL is observed, which may be because the corresponding switching field from WAL to WL is larger than 50 T. Hall trace of sample 2, as shown in Fig. S2, reveals the sheet carrier density $n_s \sim 2.12 \times 10^{13}$ cm$^{-2}$ and mobility $\mu \sim 245$ cm$^2$/(V·s) at 4.2 K. The tiny differences of transport properties between sample 1 and sample 2 may be caused by different sample qualities.

III. Control experiments

To have a comprehensive understanding of the WAL-WL crossover, control experiments for sample 1 are operated. As shown in Fig. S3, the MR behaviors, even the absolute values of the resistance, measured by 3 μA and 15 μA excitation current are identical, so are the MR data under opposite current direction. Current intensity and direction seem to count little for the MR behaviors.

We study the MR along [110] and [-110] directions by standard four-probe and quasi-four-probe method, respectively, thus the differences between the two measurement configurations are further studied at 77 K with the same voltage distance in sample 1, as shown in the insets of Fig. S4. The MR change obtained by quasi-four-probe measurement is about 50% smaller than that measured by standard four-probe method, which may be because the former can introduce contact resistance and then reduce the negative MR amplitude.

**Figure legends**

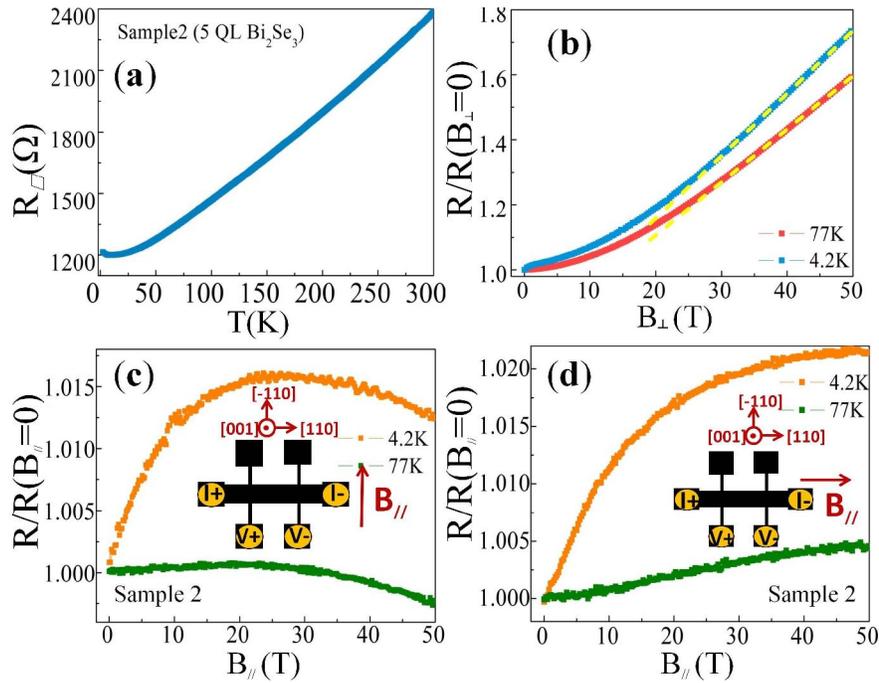

**Figure S1 Transport properties of Sample 2.** (a) Sheet resistance vs. temperature. Normalized magneto-resistance (b) in perpendicular field and at the configuration of (c) $B_{//} \perp I$ and (d) $B_{//} // I$. The black lines in (b) are guides to the eyes. The insets show the corresponding measurement schemas. (110) and (-110) mean two different crystal directions along the film plane. (001) means the crystal direction perpendicular to the film plane.

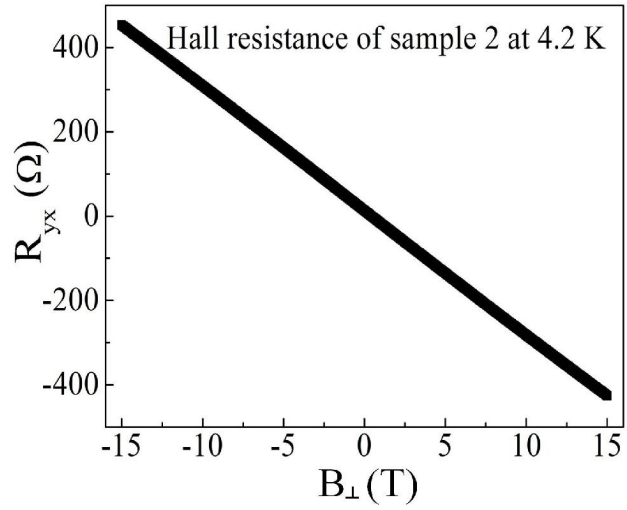

**Figure S2** Hall resistance of sample 2 versus the perpendicular magnetic field at 4.2 K.

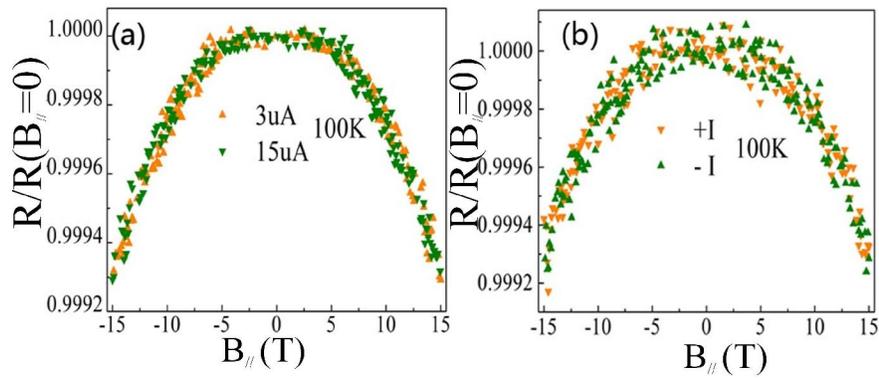

**Figure S3** Normalized MR of sample 1 along [110] direction measured in parallel magnetic field with the excitation current of (a) 3 μA and 15 μA and (b) opposite directions.

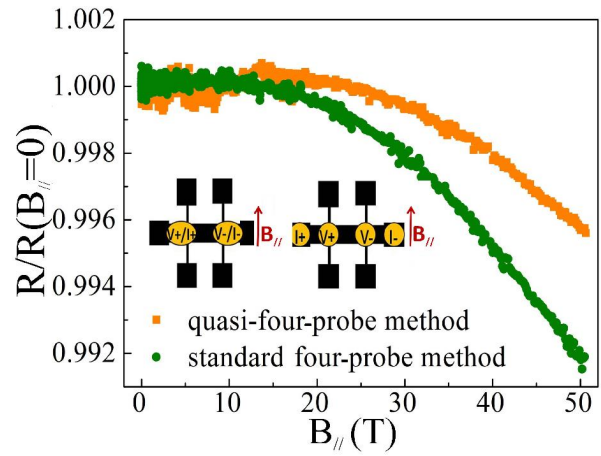

**Figure S4** Differences of the measured parallel field MR results at 77 K between by standard four-probe and by quasi-four-probe method. The left and right insets show the measurement structures of the quasi-four-probe and standard four-probe method, respectively. The voltage distances between V+ and V- for two methods are same.

[1] B. L. Altshuler, A. G. Aronov, D. E. Khmelnitskii, and A. I. Larkin, *Quantum Theory of Solids*, edited by I. M. Lifshitz (Mir, Moscow, 1982), P. 130.

[2] V. K. Dugaev and D. E. Khmelnitskii, Sov. Phys. JETP 59, 1038 (1984).